\documentclass[prl,aps]{revtex4}

\begin{document}
\title{Estimating the density scaling exponent of viscous liquids from specific heat and bulk modulus data}

\author{Ulf R. Pedersen, Tina Hecksher, Bo Jakobsen, Thomas B. Schr{\o}der, Nicoletta Gnan, Nicholas P. Bailey, Jeppe C. Dyre}

\affiliation{DNRF Centre ``Glass and Time,'' IMFUFA, Department of Sciences, Systems and Models,
Roskilde University, Postbox 260, DK-4000 Roskilde, Denmark}

\begin{abstract}
It was recently shown by computer simulations that a large class of liquids exhibits strong correlations in their thermal fluctuations of virial and potential energy [Pedersen {\it et al.}, Phys. Rev. Lett. {\bf 100}, 015701 (2008)]. Among organic liquids the class of strongly correlating liquids includes van der Waals liquids, but excludes ionic and hydrogen-bonding liquids. The present note focuses on the density scaling of strongly correlating liquids, i.e., the fact their relaxation time $\tau$ at different densities $\rho$ and temperatures $T$ collapses to a master curve according to the expression $\tau\propto F[\rho^\gamma/T]$ [Schr{\o}der {\it et al.}, arXiv:0803.2199]. We here show how to calculate the exponent $\gamma$ from bulk modulus and specific heat data, either measured as functions of frequency in the metastable liquid or extrapolated from the glass and liquid phases to a common temperature (close to the glass transition temperature). Thus an exponent defined from the response to highly nonlinear parameter changes may be determined from linear response measurements.
\end{abstract}

\maketitle

\subsection{Introduction}

A liquid is termed ``strongly correlating'' if its viral ($W$) and potential energy ($U$) equilibrium fluctuations correlate better than 90\% \cite{Pedersen2008PhysRev,bailey2008I,bailey2008II} at constant volume and temperature,
\begin{equation}
\Delta W(t)\cong \gamma \Delta U(t).
\end{equation}
where $\Delta W(t)=W(t)-\langle W\rangle$, $\Delta U(t)=U(t)-\langle U\rangle$. The ``slope'' is defined as 
\begin{equation}\label{slopedef}
	\gamma\equiv\sqrt{\frac{\langle(\Delta W)^2\rangle}{\langle(\Delta U)^2\rangle}}
\end{equation}
and the correlation coefficient is defined as
\begin{equation}\label{Rdef}
	R\equiv\frac{\langle\Delta W \Delta U\rangle}{\sqrt{\langle(\Delta W)^2\rangle\langle(\Delta U)^2\rangle}}.
\end{equation}
We have previously shown that the fluctuations of such liquids are well described by those generated by soft-sphere potentials (inverse power law potentials) \cite{bailey2008I,bailey2008II}. Strongly correlating liquids are approximate single-parameter liquids \cite{Pedersen2008PhysRevE,ellegaard2007,bailey2008Jphys}. Moreover the density scaling exponent is one third of the exponent of the approximate inverse power law potentials \cite{schroder2000DensArxiv,coslovich2009}. Recall that 
density (thermodynamic) scaling applies whenever the relaxation time $\tau$ at different densities $\rho$ and temperatures $T$ collapse to a master curve according to the expression $\tau\propto F[\rho^\gamma/T]$. It is generally reported now that density scaling applies for van der Waals liquids, but e.g. not for hydrogen-bonding liquids. This is consistent with our finding that the class of strongly correlating liquids includes van der Waals and metallic liquids, but excludes covalent, ionic, or hydrogen-bonding liquids -- the latter three classes of liquids have competing interactions that spoil the $WU$ correlation \cite{Pedersen2008PhysRev,bailey2008I,bailey2008II}.

\subsection{Fluctuation expressions}

Consider a viscous liquid with slow structural relaxation, i.e., with a relaxation time $\tau$ that is much larger than one picosecond. The fluctuation-dissipation (FD) theorem for the frequency-dependent specific heat per unit volume, $c_V(\omega)$, is given \cite{nielsen1996} by

\begin{equation}\label{fd}
k_BT^2V\,c_V(\omega)\,=\,
\langle(\Delta E)^2\rangle-i\omega\int_0^\infty\langle\Delta E(0)\Delta E(t)\rangle e^{-i\omega t}dt\,.
\end{equation}
Relaxation takes place over a limited range of frequencies  -- typically 3-5 decades. By subtracting the responses at high (``$\infty$'') and low (``$0$'') frequencies well outside the relaxation frequency range, it follows that if $t_s$ is a time much shorter than those of the relaxations, but much longer than one picosecond, then

\begin{equation}\label{fd_sub}
k_BT^2V\,\left(c_V(0)-c_V(\infty)\right)\,=\,
\langle\Delta E(0)\Delta E(t_s)\rangle\,.
\end{equation}
Write the energy as potential plus kinetic energy, $E=U+K$. Because the kinetic energy in the NVT ensemble fluctuates fast compared to $t_s$ one has $\langle\Delta E(0)\Delta E(t_s)\rangle\cong\langle\Delta U(0)\Delta U(t_s)\rangle$. Thus,

\begin{equation}\label{dcveq}
k_BT^2V\,\left(c_V(0)-c_V(\infty)\right)\,=\,
\langle(\Delta U)^2\rangle^\textrm{slow}\,
\end{equation}
where $\langle(\Delta U)^2\rangle^\textrm{slow}=\langle\Delta U(0)\Delta U(t_s)\rangle$ is the slow part of the potential energy fluctuations, i.e., slow compared to the picosecond time scale.

Since the low-frequency limit gives the ordinary (dc) liquid specific heat and the high-frequency limit gives the ``glassy'' specific heat corresponding to perturbations that probe a frozen structure, this result may be written 

\begin{equation} \label{CUU}
\langle(\Delta U)^2\rangle^\textrm{slow}
=k_BT^2V(c_V^\textrm{liquid}-c_V^\textrm{solid})\,.
\end{equation}
Similarly one finds for the virial fluctuations (where $K_T$ is the isothermal bulk modulus)

\begin{equation}
	\langle(\Delta W)^2\rangle^\textrm{slow}\,=\,-\,k_BTV (K_T^\textrm{liquid}-K_T^\textrm{solid})
\end{equation}
and for the virial / potential energy correlation (where $\beta$ is the pressure coefficient)
\begin{equation}  \label{CWU}
	\langle\Delta W \Delta U\rangle^\textrm{slow}=k_BT^2V(\beta_V^\textrm{liquid}-\beta_V^\textrm{solid}) \,.
\end{equation}

\subsection{Calculating $R$ and $\gamma$ from data}

The $WU$ correlation coefficient (Eq. \ref{Rdef}) on the $t_s$ timescale can now be expressed in terms of experimental linear response quantities as follows:
\begin{equation}\label{Rresp}
	R=\frac{ (\beta_V^\textrm{liquid}-\beta_V^\textrm{solid}) }
{\sqrt{-(K_T^\textrm{liquid}-K_T^\textrm{solid})(c_V^\textrm{liquid}-c_V^\textrm{solid})/T}}\,.
\end{equation}
Similarly the slope (Eq. \ref{slopedef}) can be calculated as
\begin{equation}\label{gammaresp}
	\gamma=\sqrt{\frac{ -(K_T^\textrm{liquid}-K_T^\textrm{solid}) }{ T(c_V^\textrm{liquid}-c_V^\textrm{solid})  }}.
\end{equation}

\subsection{Calculating $\gamma$ for the commercial silicone oil DC704}

The slope $\gamma$ may be estimated from the high- and low-frequency limits of $c_p(\omega)$ and $K_S(\omega)$ (and a rough estimate of $\alpha_p$) by proceeding as follows. Recall the identities

\begin{equation}
c_V=c_p-T\alpha_p^2K_T
\end{equation}
and

\begin{equation}
K_T=K_S\frac{c_V}{c_p}.
\end{equation}
Combining these we get

\begin{equation}
c_V=c_p\left(1+\frac{T\alpha_p^2K_S}{c_p}\right)^{-1},
\end{equation}
and

\begin{equation}
K_T=K_S\left(1+\frac{T\alpha_p^2K_S}{c_p}\right)^{-1}.
\end{equation}
The value $\gamma=6$ is calculated from these expressions using our unpublished linear response data (table \ref{tab_DC704}). Unfortunately no experimental density scaling $\gamma$'s are available for this liquid to compare to.

\begin{table}
\begin{center}
\caption{$\gamma$ of DC704 data of from the high and low frequency limit of $c_p(\omega)$ and $K_S(\omega)$ \cite{tinaBo2009}. $^a$ $\alpha_p^\textrm{solid}$ is estimated as a typical value of solids.}\label{tab_DC704}
\begin{tabular}{lcc}
\hline
$T$ [K] 			 &  214  \\
\hline
$c_p^\textrm{liquid}$ [$10^6$ J/(K m$^3$)] & 1.40  \\
$c_p^\textrm{solid}$ [$10^6$ J/(K m$^3$)] & 1.05  \\
$K_S^\textrm{liquid}$ [$10^{9}$ Pa]	&   3.6 \\
$K_S^\textrm{solid}$ [$10^{9}$ Pa]	&   5.1 \\
$\alpha_p^\textrm{liquid}$ [$10^{-3}$ K$^{-1}$] &   0.5 \\
$\alpha_p^\textrm{solid}$ [$10^{-3}$ K$^{-1}$]	&   0.1$^a$ \\
\hline
$T(\alpha_p^\textrm{liquid})^2K_S^\textrm{liquid}/c_p^\textrm{liquid}$ & 0.13 \\
$T(\alpha_p^\textrm{solid})^2K_S^\textrm{solid}/c_p^\textrm{solid}$ & 0.01\\
$c_V^\textrm{liquid}$ [$10^6$ J/(K m$^3$)] & 1.24 \\
$c_V^\textrm{solid}$ [$10^6$ J/(K m$^3$)] & 1.04 \\
$K_T^\textrm{liquid}$ [$10^{9}$ Pa]	&   3.1  \\
$K_T^\textrm{solid}$ [$10^{9}$ Pa]	&   5.0  \\
\hline
$\gamma$ (Eq. \ref{gammaresp}) & 6 \\
\hline
\end{tabular}
\end{center}
\end{table}

\subsection{OTP/OPP mixture and pure OTP}

The high- and low-frequency limits of the dynamic response can be estimated by extrapolation of static response functions of the glass and liquid phases to a temperature close to the glass transition temperature $T_g$.
Table \ref{tab_otpopp} lists extrapolated values of $\kappa_T$, $c_p$ and $\alpha_p$ for a mixture of o-terphenyl (OTP) and o-phenylphenol (OPP), Table \ref{tab_otp} lists values for pure OTP.

From the natural response functions of the constant $pT$ ensemble it is straightforward to calculate the natural response functions of the constant $VT$ ensemble:
\begin{equation}
	K_T=1/\kappa_T,
\end{equation}
\begin{equation}
	c_V=c_p-T\alpha_p^2K_T,
\end{equation}
and
\begin{equation}
	\beta_V=\alpha_p K_T.
\end{equation}
Using these equations we arrive at the numbers in tables \ref{tab_otpopp} and \ref{tab_otp}. 

\begin{table}
\begin{center}
\caption{$\gamma$ and $R$ of OTP-OPP calculated from data on Figure 4 of Ref. \cite{takahara1999} }\label{tab_otpopp}
\begin{tabular}{lcc}
\hline
$T_g$ [K] 			 &  233.7  \\
$V_g$ [$10^{-6}$m$^3$/mol]	 &  203.9  \\
\hline
$C_p^\textrm{liquid}$ [J/(K mol)] & 364  \\
$C_p^\textrm{solid}$ [J/(K mol)]	 & 236 \\
$\kappa_T^\textrm{liquid}$ [$10^{-9}$ Pa$^{-1}$] & 0.35 \\
$\kappa_T^\textrm{solid}$ [$10^{-9}$ Pa$^{-1}$]  & 0.19 \\
$\alpha_p^\textrm{liquid}$ [$10^{-3}$ K$^{-1}$] & 0.74 \\
$\alpha_p^\textrm{solid}$ [$10^{-3}$ K$^{-1}$]	 & 0.17 \\
\hline
$C_V^\textrm{liquid}$ [J/(K mol)]	&  284 \\
$C_V^\textrm{solid}$ [J/(K mol)]	&  229 \\
$K_T^\textrm{liquid}$ [$10^{9}$ Pa]	&  2.9  \\
$K_T^\textrm{solid}$ [$10^{9}$ Pa]	&  5.2 \\
$\beta_V^\textrm{liquid}$ [$10^6$ Pa/K] &  2.1 \\
$\beta_V^\textrm{solid}$ [$10^6$ Pa/K]	&  0.9 \\
\hline
$R$ (Eq. \ref{Rresp})	&   0.8 \\
$\gamma$ (Eq. \ref{gammaresp}) &  6.0 \\
\hline
$\gamma^\textrm{scale}$ (Ref. \cite{roland2005}) & 6.2 \\
\hline
\end{tabular}
\end{center}
\end{table}

\begin{table}
\begin{center}
\caption{$\gamma$ and $R$ for OTP. $C_p$ values are from Ref. \cite{chang1972}. $K_T=\beta_V/\alpha_p$ is calculated from $\alpha_p$ and $\beta_V$ values from Ref. \cite{naoki1989}.}\label{tab_otp}
\begin{tabular}{lcc}
\hline
$T_g$ [K] 			 &  244.5  \\
$V_g$ [$10^{-6}$m$^3$/mol]	 &  206.1  \\
\hline
$C_p^\textrm{liquid}$ [J/(K mol)] & 336  \\
$C_p^\textrm{solid}$ [J/(K mol)] & 228 \\
$\alpha_p^\textrm{liquid}$ [$10^{-3}$ K$^{-1}$] & 0.71 \\
$\alpha_p^\textrm{solid}$ [$10^{-3}$ K$^{-1}$]	 & 0.32 \\
$\beta_V^\textrm{liquid}$ [$10^6$ Pa/K]	& 1.56 \\
$\beta_V^\textrm{solid}$ [$10^6$ Pa/K]	& 1.16  \\
\hline
$C_V^\textrm{liquid}$ [J/(K mol)] & 280 \\
$C_V^\textrm{solid}$ [J/(K mol)] & 209 \\
$K_T^\textrm{liquid}$ [$10^{9}$ Pa] & 2.2 \\
$K_T^\textrm{solid}$ [$10^{9}$ Pa] & 3.6 \\
\hline
$R$ (Eq. \ref{Rresp})		&  0.3  \\
$\gamma$ (Eq. \ref{gammaresp}) &  4.1 \\
\hline
$\gamma^\textrm{scale}$ (Ref. \cite{roland2005}) & 4.0 \\
\hline
\end{tabular}
\end{center}
\end{table}

\subsection{Summary}

We have shown that it is possible to calculate the density scaling exponent $\gamma$ from linear response measurements of specific heat and bulk modulus data. 
There are two ways to do this: 
Either by measuring broad-range frequency-dependent linear responses in the equilibrium metastable liquid phase or by extrapolations as done when evaluating the Prigogine-Defay ratio \citep{prigogine1954}. 
Using the first method for DC704 we find $\gamma=6$. 
To the best of our knowledge there are yet no density scaling data for this liquid. 
Using the second method for the OTP-OPP mixture we find $\gamma=6.0$ which compares favorably to the density scaling $\gamma^\textrm{scale}=6.2$ \cite{roland2005}; 
similarly we find $\gamma=4.1$ for pure OTP that compares favorably to the density scaling $\gamma^\textrm{scale}=4.0$ \cite{roland2005}. 
This good agreement may well be fortuitous given the uncertainties associated with our $\gamma$ estimates. 
Nevertheless these preliminary findings suggest that for strongly correlating liquids (``single-parameter liquids'') the density scaling exponent 
-- which refers to highly nonlinear parameter changes -- may be determined from linear response measurements. 
This is consistent with a general hypothesis of ours that strongly correlating liquids have simpler physics than liquids in general.

\bibliographystyle{plain}
\bibliography{../sclExp}

\end{document}